\documentstyle[12pt,color,epsfig,subfigure,ulem]{article}
\def\baselinestretch{1.3}
\advance\textheight by \topskip
\newcommand{\comment}[1]{}
\begin{document}
%\linenumbers
\tolerance=100000
\thispagestyle{empty}
\setcounter{page}{0}
\topmargin -0.1in
\headsep 30pt
\footskip 40pt
\oddsidemargin 12pt
\evensidemargin -16pt
\textheight 8.5in
\textwidth 6.5in
\parindent 20pt
 
\def\baselinestretch{1.5}
%%\def\baselinestretch{1.1}
%-------- General character and macros start -----------------
\newcommand{\newc}{\newcommand}
\def\preprint{{preprint}}
\def\Ord{\lower .7ex\hbox{$\;\stackrel{\textstyle <}{\sim}\;$}}
\def\OOrd{\lower .7ex\hbox{$\;\stackrel{\textstyle >}{\sim}\;$}}
% --- calligraphic stuff ---
\def\cO#1{{\cal{O}}\left(#1\right)}
%\DeclareRobustCommand*\cal{\@fontswitch\relax\mathcal}
\newc{\order}{{\cal O}}
\def\lag             {{\cal L}}
\def\Lag             {{\cal L}}
\def\lum             {{\cal L}}
\def\R               {{\cal R}}
\def\Rsq             {{\cal R}^{\sq}}
\def\Rst             {{\cal R}^{\st}}
\def\Rsb             {{\cal R}^{\sb}}
\def\M               {{\cal M}}
\def\Oas             {{\cal O}(\alpha_{s})}
\def\Vcal            {{\cal V}}
\def\Wcal            {{\cal W}}
\newc{\be}{\begin{equation}}
\newc{\ee}{\end{equation}}
\newc{\br}{\begin{eqnarray}}
\newc{\er}{\end{eqnarray}}
\newc{\ba}{\begin{array}}
\newc{\ea}{\end{array}}
\newc{\bi}{\begin{itemize}}
\newc{\ei}{\end{itemize}}
\newc{\bn}{\begin{enumerate}}
\newc{\en}{\end{enumerate}}
\newc{\bc}{\begin{center}}
\newc{\ec}{\end{center}}
\newc{\ul}{\underline}
\newc{\ol}{\overline}
\newc{\ra}{\rightarrow}
%\newc{\to}{\ra}
\newc{\lra}{\longrightarrow}
\newc{\wt}{\widetilde}
\newc{\til}{\tilde}
\def\kr              {^{\dagger}}
\newc{\wh}{\widehat}
\newc{\ti}{\times}
\newc{\Dir}{\kern -6.4pt\Big{/}}
\newc{\Dirin}{\kern -10.4pt\Big{/}\kern 4.4pt}
\newc{\DDir}{\kern -10.6pt\Big{/}}
\newc{\DGir}{\kern -6.0pt\Big{/}}
\newc{\sig}{\sigma}
\newc{\sigmalstop}{\sig_{\lstoppair}}
\newc{\Sig}{\Sigma}  %%\def\Sig{\Sigma} this is the def style
\newc{\del}{\delta}
\newc{\Del}{\Delta}
\newc{\lam}{\lambda}
\newc{\Lam}{\Lambda}
\newc{\gam}{\gamma}
\newc{\Gam}{\Gamma}
\newc{\eps}{\epsilon}
\newc{\Eps}{\Epsilon}
\newc{\kap}{\kappa}
\newc{\Kap}{\Kappa}
\newc{\modulus}[1]{\left| #1 \right|}
\newc{\eq}[1]{(\ref{eq:#1})}
\newc{\eqs}[2]{(\ref{eq:#1},\ref{eq:#2})}
\newc{\etal}{{\it et al.}\ }
\newc{\ibid}{{\it ibid}.}
\newc{\ibidem}{{\it ibidem}.}
\newc{\eg}{{\it e.g.}\ }
\newc{\ie}{{\it i.e.}\ }
\def \viz{\emph{viz.}}
\def \etc{\emph{etc. }}
\newc{\nonum}{\nonumber}
\newc{\lab}[1]{\label{eq:#1}}
\newc{\dpr}[2]{({#1}\cdot{#2})}
\newc{\lt}{\stackrel{<}}
\newc{\gt}{\stackrel{>}}
\newc{\lsimeq}{\stackrel{<}{\sim}}
\newc{\gsimeq}{\stackrel{>}{\sim}}
\def\lsim{\buildrel{\scriptscriptstyle <}\over{\scriptscriptstyle\sim}}
\def\gsim{\buildrel{\scriptscriptstyle >}\over{\scriptscriptstyle\sim}}
%\newcommand{\lsim}{\raisebox{-0.13cm}{~\shortstack{$<$ \\[-0.07cm] $\sim$}}~}
%\newcommand{\gsim}{\raisebox{-0.13cm}{~\shortstack{$>$ \\[-0.07cm] $\sim$}}~}
%sim single and approx double sim
\def\lapp{\mathrel{\rlap{\raise.5ex\hbox{$<$}}
                    {\lower.5ex\hbox{$\sim$}}}}
\def\gapp{\mathrel{\rlap{\raise.5ex\hbox{$>$}}
                    {\lower.5ex\hbox{$\sim$}}}}
\newc{\half}{\frac{1}{2}}
\newcommand {\nnc}        {{\overline{\mathrm N}_{95}}}
\newcommand {\dm}         {\Delta m}
\newcommand {\dM}         {\Delta M}
\def\bra{\langle}
\def\ket{\rangle}
\def\cO#1{{\cal{O}}\left(#1\right)}
\def \DM{{\Delta{m}}}
\newc{\bQ}{\ol{Q}}
\newc{\dota}{\dot{\alpha }}
\newc{\dotb}{\dot{\beta }}
\newc{\dotd}{\dot{\delta }}
\newc{\nindnt}{\noindent}

% Def. for small \fracs:
\newcommand{\medf}[2] {{\footnotesize{\frac{#1}{#2}} }}
\newcommand{\smaf}[2] {{\textstyle \frac{#1}{#2} }}
\def\onesq            {{\textstyle \frac{1}{\sqrt{2}} }}
\def\onehf            {{\textstyle \frac{1}{2} }}
\def\oneth            {{\textstyle \frac{1}{3} }}
\def\twoth            {{\textstyle \frac{2}{3} }}
\def\onefo            {{\textstyle \frac{1}{4} }}
\def\forth            {{\textstyle \frac{4}{3} }}

\newc{\matth}{\mathsurround=0pt}
\def\ML{\ifmmode{{\mathaccent"7E M}_L}
             \else{${\mathaccent"7E M}_L$}\fi}
\def\MR{\ifmmode{{\mathaccent"7E M}_R}
             \else{${\mathaccent"7E M}_R$}\fi}
\newcommand{\s}{\\ \vspace*{-3mm} }

\def \ud { {1 \over 2} }
\def \ut { {1 \over 3} }
\def \td { {3 \over 2} }
\newc{\mr}{\mathrm}
\def\dh {\partial }
\def \cs { cross-section }
\def \css { cross-sections }
\def \cm { centre of mass }
\def \cms { centre of mass energy }
\def \cc { coupling constant }
\def \ccs {coupling constants }
\def \gc {gauge coupling }
\def \gcc {gauge coupling constant }
\def \gccs {gauge coupling constants }
\def \yc {Yukawa coupling }
\def \ycc {Yukawa coupling constant }
\def \pp {{parameter }}
\def \pps {{parameters }} % spdas {{ or { which is correct 
% if not take from alephstop
\def \ps {parameter space }
\def \pss {parameter spaces }
\def \vv {vice versa }

\newc{\siminf}{\mbox{$_{\sim}$ {\small {\hspace{-1.em}{$<$}}}    }}
\newc{\simsup}{\mbox{$_{\sim}$ {\small {\hspace{-1.em}{$>$}}}    }}

%------------------------------------------------------------

% SM notations
\newc {\Zboson}{{\mathrm Z}^{0}}
\newc{\thetaw}{\theta_W}
\newc{\mbot}{{m_b}}
\newc{\mtop}{{m_t}}
\newc{\sm}{${\cal {SM}}$}
\newc{\as}{\alpha_s}
\newc{\aem}{\alpha_{em}}
\def \PI{{\pi^{\pm}}}
\newc{\ppbar}{\mbox{$p\ol{p}$}}
\newc{\bbbar}{\mbox{$b\ol{b}$}}
\newc{\ccbar}{\mbox{$c\ol{c}$}}
\newc{\ttbar}{\mbox{$t\ol{t}$}}
\newc{\eebar}{\mbox{$e\ol{e}$}}
\newc{\zzero}{\mbox{$Z^0$}}
\def \gamz{\Gam_Z}
\newc{\wplus}{\mbox{$W^+$}}
\newc{\wminus}{\mbox{$W^-$}}
\newc{\ellp}{\ell^+}
\newc{\ellm}{\ell^-}
\newc{\elp}{\mbox{$e^+$}}
\newc{\elm}{\mbox{$e^-$}}
\newc{\elpm}{\mbox{$e^{\pm}$}}
\newc{\qbar}     {\mbox{$\ol{q}$}}
%%\newc{\lp}{\mbox{$e^+$}}
%%\newc{\lm}{\mbox{$e^-$}}
%%\newc{\lpm}{\mbox{$e^{\pm}$}}
\def \ewgroup{SU(2)_L \otimes U(1)_Y}
\def \smgroup{SU(3)_C \otimes SU(2)_L \otimes U(1)_Y}
\def \smcolorem{SU(3)_C \otimes U(1)_{em}}
%---------------------------------------------------------

%SUSY notations
\def \SSM  {Supersymmetric Standard Model}
\def \poincare{Poincare$\acute{e}$}
\def \superspace{\emph{superspace}}
\def \sfs{\emph{superfields}}
\def \superpot{\emph{superpotential}}
\def \csf{\emph{chiral superfield}}
\def \csfs{\emph{chiral superfields}}
\def \vsf{\emph{vector superfield }}
\def \vsfs{\emph{vector superfields}}
\newc{\Ebar}{{\bar E}}
\newc{\Dbar}{{\bar D}}
\newc{\Ubar}{{\bar U}}
\newc{\susy}{{{SUSY}}}
\newc{\msusy}{{{M_{SUSY}}}}
%----------------------------------------------------

%Gauginos
\def\photino{\ifmmode{\mathaccent"7E \gam}\else{$\mathaccent"7E \gam$}\fi}
\def\taugluino{\ifmmode{\tau_{\mathaccent"7E g}}
             \else{$\tau_{\mathaccent"7E g}$}\fi}
\def\mphotino{\ifmmode{m_{\mathaccent"7E \gam}}
             \else{$m_{\mathaccent"7E \gam}$}\fi}
\newc{\gl}   {\mbox{$\wt{g}$}}
\newc{\mgl}  {\mbox{$m_{\gl}$}}
%%\newc{\gl}{\wt g}
%%\newc{\mgl}{m_{\gl}}
%%\def\gluino{\ifmmode{\mathaccent"7E g}\else{$\mathaccent"7E g$}\fi}
%%\def\mgluino{\ifmmode{m_{\mathaccent"7E g}}
%%             \else{$m_{\mathaccent"7E g}$}\fi}
%----------------------------------------------------
% Chargino
\def \charginopm{{\wt\chi}^{\pm}}
\def \mcharginopm{m_{\charginopm}}
\def \mchpmmin {\mcharginopm^{min}}
\def \chonep {{\wt\chi_1^+}}
\def \chone {{\wt\chi_1}}
\def \ch2p {{\wt\chi_2^+}}
\def \chonem {{\wt\chi_1^-}}
\def \ch2m {{\wt\chi_2^-}}
\def \chplus {{\wt\chi^+}}
\def \chminus {{\wt\chi^-}}
\def \chonip{{\wt\chi_i}^{+}}
\def \chonim{{\wt\chi_i}^{-}}
\def \chonipm{{\wt\chi_i}^{\pm}}
\def \chonjp{{\wt\chi_j}^{+}}
\def \chonjm{{\wt\chi_j}^{-}}
\def \chonjpm{{\wt\chi_j}^{\pm}}
\def \chonepm{{\wt\chi_1}^{\pm}}
\def \chonemp{{\wt\chi_1}^{\mp}}
\def \mchonepm{m_{\chonepm}}
\def \mchonemp{m_{\chonemp}}
\def \chtwopm{{\wt\chi_2}^{\pm}}
\def \mchtwopm{m_{\chtwopm}}
\newc{\dmchi}{\Delta m_{\wt\chi}}

%-----------------------------------------------------------------------
% Neutralino

\def \vlsp{\emph{VLSP}}
\def \lspi{\wt\chi_i^0}
\def \mlspi{m_{\lspi}}
\def \lspj{\wt\chi_j^0}
\def \mlspj{m_{\lspj}}
\def \lspone{\wt\chi_1^0}
\def \mlspone{m_{\lspone}}
\def \lsptwo{\wt\chi_2^0}
\def \mlsptwo{m_{\lsptwo}}
\def \lspthree{\wt\chi_3^0}
\def \mlspthree{m_{\lspthree}}
\def \lspfour{\wt\chi_4^0}
\def \mlspfour{m_{\lspfour}}

%-----------------------------------------------------------------------
%SLEPTONs

\newc{\sele}{\wt{\mathrm e}}
\newc{\sell}{\wt{\ell}}
\def \msell{m_{\sell}}
\def \slepone{\wt\ell_1}
\def \mslepone{m_{\slepone}}
\def \smuone{\wt\mu_1}
\def \msmuone{m_{\smuone}}
\def \stauone{\wt\tau}
\def \mstauone{m_{\stauone}}
\def \snu{\wt{\nu}}
\def \snutau{\wt{\nu}_{\tau}}
\def \msnu{m_{\snu}}
\def \msnumu{m_{\snu_{\mu}}}
\def \barsnu{\wt{\bar{\nu}}}
\def \barsnul{\barsnu_{\ell}}
\def \snul{\snu_{\ell}}
\def \mbarsnu{m_{\barsnu}}
\newc{\snue}     {\mbox{$ \wt{\nu_e}$}}
\newc{\smu}{\wt{\mu}}
\newc{\stau}{\wt{\tau}}
%%% from 9810232.tex
\newc {\nuL} {\wt{\nu}_L}
\newc {\nuR} {\wt{\nu}_R}
\newc {\snub} {\bar{\wt{\nu}}}
\newc {\eL} {\wt{e}_L}
\newc {\eR} {\wt{e}_R}
%%% from 9810232.tex
\def \slep{\wt{l}}
\def \slepl{\wt{l}_L}
\def \mslepl{m_{\slepl}}
\def \slepr{\wt{l}_R}
\def \mslepr{m_{\slepr}}
\def \stau{\wt\tau}
\def \mstau{m_{\stau}}
\def \slepton{\wt\ell}
\def \mslepton{m_{\slepton}}
\def \mlhiggs{m_{h^0}}

%----------------------------------------------------
%SQUARKs
\def \xr{X_{r}}

%\newc{\sq}   {\mbox{$\wt{q}$}}
%\newc{\msq}  {\mbox{$m_(\sq)$}}
\def \sfer{\wt{f}}
\def \msfer{m_{\sfer}}
\def \sq{\wt{q}}
\def \msq{m_{\sq}}
\def \msquleft{m_{\tilde{u_L}}}
\def \msqurht{m_{\tilde{u_R}}}
\def \sql{\wt{q}_L}
\def \msql{m_{\sql}}
\def \sqr{\wt{q}_R}
\def \msqr{m_{\sqr}}
\newc{\msqot}  {\mbox{$m_(\sq_{1,2} )$}}
\newc{\sqbar}    {\mbox{$\bar{\wt{q}}$}}
\newc{\ssb}      {\mbox{$\squark\ol{\squark}$}}
\newc {\qL} {\wt{q}_L}
\newc {\qR} {\wt{q}_R}
\newc {\uL} {\wt{u}_L}
\newc {\uR} {\wt{u}_R}
\def \ul{\wt{u}_L}
\def \mul{m_{\ul}}
\newc {\dL} {\wt{d}_L}
\newc {\dR} {\wt{d}_R}
\newc {\cL} {\wt{c}_L}
\newc {\cR} {\wt{c}_R}
\newc {\sL} {\wt{s}_L}
\newc {\sR} {\wt{s}_R}
\newc {\tL} {\wt{t}_L}
\newc {\tR} {\wt{t}_R}
\newc {\stb} {\ol{\wt{t}}_1}
\newc {\sbot} {\wt{b}_1}
\newc {\msbot} {m_{\sbot}}
\newc {\sbotb} {\ol{\wt{b}}_1}
\newc {\bL} {\wt{b}_L}
\newc {\bR} {\wt{b}_R}
\def \mul{m_{\wt{u}_L}}
\def \mur{m_{\wt{u}_R}}
\def \mdl{m_{\wt{d}_L}}
\def \mdr{m_{\wt{d}_R}}
\def \mcl{m_{\wt{c}_L}}
\def \charml{\wt{c}_L}
\def \mcr{m_{\wt{c}_R}}
\newc{\csquark}  {\mbox{$\wt{c}$}}
\newc{\csquarkl} {\mbox{$\wt{c}_L$}}
\newc{\mcsl}     {\mbox{$m(\csquarkl)$}}
\def \msl{m_{\wt{s}_L}}
\def \msr{m_{\wt{s}_R}}
\def \mbl{m_{\wt{b}_L}}
\def \mbr{m_{\wt{b}_R}}
\def \mtl{m_{\wt{t}_L}}
\def \mtr{m_{\wt{t}_R}}
\def \st{\wt{t}}
\def \mst{m_{\st}}
\newc {\stopl}         {\wt{\mathrm{t}}_{\mathrm L}}
\newc {\stopr}         {\wt{\mathrm{t}}_{\mathrm R}}
\newc {\stoppair}      {\wt{\mathrm{t}}_{1}
\bar{\wt{\mathrm{t}}}_{1}}
\def \lstop{\wt{t}_{1}}
\def \lstopbar{\lstop^*}
\def \hstop{\wt{t}_{2}}
\def \hstopbar{\hstop^*}
\def \mlstop{m_{\lstop}}
\def \mhstop{m_{\hstop}}
\def \lstoppair{\lstop\lstop^*}
\def \hstoppair{\hstop\hstop^*}
\newc{\tsquark}  {\mbox{$\wt{t}$}}
\newc{\ttb}      {\mbox{$\tsquark\ol{\tsquark}$}}
\newc{\ttbone}   {\mbox{$\tsquark_1\ol{\tsquark}_1$}}
% top squark related
\def \tsq {top squark }
\def \tsqs {top squarks }
\def \tsql {ligtest top squark }
\def \tsqh {heaviest top squark }
\newc{\mix}{\theta_{\wt t}}
\newc{\cost}{\cos{\theta_{\wt t}}}
\newc{\sint}{\sin{\theta_{\wt t}}}
\newc{\costloop}{\cos{\theta_{\wt t_{loop}}}}
%-----------------------------------------------------------------------
\def \lsbot{\wt{b}_{1}}
\def \lsbotbar{\lsbot^*}
\def \hsbot{\wt{b}_{2}}
\def \hsbotbar{\hsbot^*}
\def \mlsbot{m_{\lsbot}}
\def \mhsbot{m_{\hsbot}}
\def \lsbotpair{\lsbot\lsbot^*}
\def \hsbotpair{\hsbot\hsbot^*}
\newc{\mixsbot}{\theta_{\wt b}}

%-----------------------------------------------------------------------
%HIGGS sector
\def \mhone{m_{h_1}}
\def \hup{{H_u}}
\def \hdn{{H_d}}
\newc{\tb}{\tan\beta}
\newc{\cb}{\cot\beta}
\newc{\vev}[1]{{\left\langle #1\right\rangle}}

%-----------------------------------------------------------------------
%SOFT BRAEAKING TERMS
\def \abot{A_{b}}
\def \atop{A_{t}}
\def \atau{A_{\tau}}
\newc{\mhalf}{m_{1/2}}
\newc{\mzero} {\mbox{$m_0$}}
\newc{\azero} {\mbox{$A_0$}}

%--------------------------------------------------------------
% RPV related  stuff
\newc{\lb}{\lam}
\newc{\lbp}{\lam^{\prime}}
\newc{\lbpp}{\lam^{\prime\prime}}
\newc{\rpv}{{\not \!\! R_p}}
\newc{\rpvm}{{\not  R_p}}
%%%\newc{\rpv}{\not R_p}
%%\newc{\rpv}{{\not \!\! R_p}}
\newc{\rp}{R_{p}}
%\newc{\rpmssm}{{$R_p$-MSSM}\ }
%\newc{\rpvmssm}{$\rpv$-MSSM}
\newc{\rpmssm}{{RPC MSSM}}
\newc{\rpvmssm}{{RPV MSSM}}

%---------------------------------------------------------------
% Collider related stuffs

\newc{\sbyb}{S/$\sqrt B$}
\newc{\pelp}{\mbox{$e^+$}}
\newc{\pelm}{\mbox{$e^-$}}
\newc{\pelpm}{\mbox{$e^{\pm}$}}
\newc{\epem}{\mbox{$e^+e^-$}}
\newc{\lplm}{\mbox{$\ell^+\ell^-$}}
\def \branch{\emph{BR}}
\def \branche{\branch(\lstop\ra be^{+}\nu_e \lspone)\ti \branch(\lstop^{*}\ra \bar{b}q\bar{q^{\prime}}\lspone)}
\def \branchmu{\branch(\lstop\ra b\mu^{+}\nu_{\mu} \lspone)\ti \branch(\lstop^{*}\ra \bar{b}q\bar{q^{\prime}}\lspone)}
\def\Ecm{\ifmmode{E_{\mathrm{cm}}}\else{$E_{\mathrm{cm}}$}\fi}
\newc{\rts}{\sqrt{s}}
\newc{\rtshat}{\sqrt{\hat s}}
\newc{\gev}{\,GeV}
\newc{\mev}{~{\rm MeV}}
\newc{\tev}  {\mbox{$\;{\rm TeV}$}}
\newc{\gevc} {\mbox{$\;{\rm GeV}/c$}}
\newc{\gevcc}{\mbox{$\;{\rm GeV}/c^2$}}
\newc{\intlum}{\mbox{${ \int {\cal L} \; dt}$}}
\newc{\call}{{\cal L}}
%%\def \miset{\not\!\!{E_T}}
%%\newc{\etmiss}{/ \hskip-7pt E_T}
%%\def \mispt{p{\!\!\!/}_T} 
\def \met  {\mbox{${E\!\!\!\!/_T}$}}
\def \cpv  {\mbox{${CP\!\!\!\!/}$}}
\newc{\ptmiss}{/ \hskip-7pt p_T}
\def \eslash{\not \! E}
\def \etslash{\not \! E_T }
\def \ptslash{\not \! p_T }
\newc{\PT}{\mbox{$p_T$}}
\newc{\ET}{\mbox{$E_T$}}
\newc{\dedx}{\mbox{${\rm d}E/{\rm d}x$}}
\newc{\ifb}{\mbox{${\rm fb}^{-1}$}}
\newc{\ipb}{\mbox{${\rm pb}^{-1}$}}
\newc{\pb}{~{\rm pb}}
\newc{\fb}{~{\rm fb}}
\newc{\ycut}{y_{\mathrm{cut}}}
\newc{\chis}{\mbox{$\chi^{2}$}}
\def \hadron{\emph{hadron}}
\def \nlc{\emph{NLC }}
\def \lhc{\emph{LHC }}
\def \cdf{\emph{CDF }}
\def\dzero{\emptyset}
\def \tevatron{\emph{Tevatron }}
\def \lep{\emph{LEP }}
\def \jets{\emph{jets }}
\def \jet(s){\emph{jet(s) }}

%-----------------------------------------------------------------------
%  Light Stop decay parameters and different modes:
\def\Crs{stroke [] 0 setdash exch hpt sub exch vpt add hpt2 vpt2 neg V currentpoint stroke 
hpt2 neg 0 R hpt2 vpt2 V stroke}
\def\loopdk{\lstop \ra c \lspone}
\def\brloopdk{\branch(\loopdk)}
\def\fourdk{\lstop \ra b \lspone  f \bar f'}
\def\brfourdk{\branch(\fourdk)}
\def\fourdklep{\lstop \ra b \lspone  \ell \nu_{\ell}}
\def\fourdkhad{\lstop \ra b \lspone  q \bar q'}
\def\brfourdklep{\branch(\fourdklep)}
\def\brfourdkhad{\branch(\fourdkhad)}
\def\twodk{\lstop \ra b \chonep}
\def\brtwodk{\branch(\twodk)}
\def\threedkslep{\lstop \ra b \wt{\ell} \nu_{\ell}}
\def\brthreedkslep{\branch(\threedkslep)}
\def\threedksnu{\lstop \ra b \wt{\nu_{\ell}} \ell }
\def\brthreedksnu{\branch(\threedksnu) }
\def\threedklsp{\lstop \ra b W \lspone }
\def\brthreedklsp{\\branch(\threedklsp) }
\def\topdk{t \ra \lstop \lspone}
\def\rpvdk{\lstop \ra e^+ d}
\def\brrpvdk{\branch(\rpvdk)}
\def\fonec{f_{11c}} 
%-----------------------------------------------------------------------
%  Scale  of Physics
\newc{\mpl}{M_{\rm Pl}}
\newc{\mgut}{M_{GUT}}
\newc{\mw}{M_{W}}
\newc{\mweak}{M_{weak}}
\newc{\mz}{M_{Z}}

%--Collabortaions ----------------------------------------
\newc{\OPALColl}   {OPAL Collaboration}
\newc{\ALEPHColl}  {ALEPH Collaboration}
\newc{\DELPHIColl} {DELPHI Collaboration}
\newc{\XLColl}     {L3 Collaboration}
\newc{\JADEColl}   {JADE Collaboration}
\newc{\CDFColl}    {CDF Collaboration}
\newc{\DXColl}     {D0 Collaboration}
\newc{\HXColl}     {H1 Collaboration}
\newc{\ZEUSColl}   {ZEUS Collaboration}
\newc{\LEPColl}    {LEP Collaboration}
\newc{\ATLASColl}  {ATLAS Collaboration}
\newc{\CMSColl}    {CMS Collaboration}
\newc{\UAColl}    {UA Collaboration}
\newc{\KAMLANDColl}{KamLAND Collaboration}
\newc{\IMBColl}    {IMB Collaboration}
\newc{\KAMIOColl}  {Kamiokande Collaboration}
\newc{\SKAMIOColl} {Super-Kamiokande Collaboration}
\newc{\SUDANTColl} {Soudan-2 Collaboration}
\newc{\MACROColl}  {MACRO Collaboration}
\newc{\GALLEXColl} {GALLEX Collaboration}
\newc{\GNOColl}    {GNO Collaboration}
\newc{\SAGEColl}  {SAGE Collaboration}
\newc{\SNOColl}  {SNO Collaboration}
\newc{\CHOOZColl}  {CHOOZ Collaboration}
\newc{\PDGColl}  {Particle Data Group Collaboration}

%--- Macros of the journal start----------------------------------------
\def\issue(#1,#2,#3){{\bf #1}, #2 (#3)}%AIP format!Vol,page(Year)
% In thesis Journal macros is exhibits to be Vol, (Year), Page
% and in the bibliography it is set like
% (vol,page,Year ) and then the below macros  
% will take care everything.
%\def\issue(#1,#2,#3){{\bf #1} (#3) #2 } % PLB format!Vol,(Year),page
\def\ASTR(#1,#2,#3){Astropart.\ Phys. \issue(#1,#2,#3)}
\def\AJ(#1,#2,#3){Astrophysical.\ Jour. \issue(#1,#2,#3)}
\def\AJS(#1,#2,#3){Astrophys.\ J.\ Suppl. \issue(#1,#2,#3)}
\def\APP(#1,#2,#3){Acta.\ Phys.\ Pol. \issue(#1,#2,#3)}
\def\JCAP(#1,#2,#3){Journal\ XX. \issue(#1,#2,#3)} %spdas
\def\SC(#1,#2,#3){Science \issue(#1,#2,#3)}
\def\PRD(#1,#2,#3){Phys.\ Rev.\ D \issue(#1,#2,#3)}
\def\PR(#1,#2,#3){Phys.\ Rev.\ \issue(#1,#2,#3)} % spdas check 
\def\PRC(#1,#2,#3){Phys.\ Rev.\ C \issue(#1,#2,#3)}
\def\NPB(#1,#2,#3){Nucl.\ Phys.\ B \issue(#1,#2,#3)}
\def\NPPS(#1,#2,#3){Nucl.\ Phys.\ Proc. \ Suppl \issue(#1,#2,#3)}
\def\NJP(#1,#2,#3){New.\ J.\ Phys. \issue(#1,#2,#3)}
\def\JP(#1,#2,#3){J.\ Phys.\issue(#1,#2,#3)}
\def\PL(#1,#2,#3){Phys.\ Lett. \issue(#1,#2,#3)}
\def\PLB(#1,#2,#3){Phys.\ Lett.\ B  \issue(#1,#2,#3)}
\def\ZP(#1,#2,#3){Z.\ Phys. \issue(#1,#2,#3)}
\def\ZPC(#1,#2,#3){Z.\ Phys.\ C  \issue(#1,#2,#3)}
\def\PREP(#1,#2,#3){Phys.\ Rep. \issue(#1,#2,#3)}
\def\PRL(#1,#2,#3){Phys.\ Rev.\ Lett. \issue(#1,#2,#3)}
\def\MPL(#1,#2,#3){Mod.\ Phys.\ Lett. \issue(#1,#2,#3)}
\def\RMP(#1,#2,#3){Rev.\ Mod.\ Phys. \issue(#1,#2,#3)}
\def\SJNP(#1,#2,#3){Sov.\ J.\ Nucl.\ Phys. \issue(#1,#2,#3)}
\def\CPC(#1,#2,#3){Comp.\ Phys.\ Comm. \issue(#1,#2,#3)}
\def\IJMPA(#1,#2,#3){Int.\ J.\ Mod. \ Phys.\ A \issue(#1,#2,#3)}
\def\MPLA(#1,#2,#3){Mod.\ Phys.\ Lett.\ A \issue(#1,#2,#3)}
\def\PTP(#1,#2,#3){Prog.\ Theor.\ Phys. \issue(#1,#2,#3)}
\def\RMP(#1,#2,#3){Rev.\ Mod.\ Phys. \issue(#1,#2,#3)}
\def\NIMA(#1,#2,#3){Nucl.\ Instrum.\ Methods \ A \issue(#1,#2,#3)}
\def\JHEP(#1,#2,#3){J.\ High\ Energy\ Phys. \issue(#1,#2,#3)}
\def\EPJC(#1,#2,#3){Eur.\ Phys.\ J.\ C \issue(#1,#2,#3)}
\def\RPP (#1,#2,#3){Rept.\ Prog.\ Phys. \issue(#1,#2,#3)}
\def\PPNP(#1,#2,#3){ Prog.\ Part.\ Nucl.\ Phys. \issue(#1,#2,#3)}
\def\JPG(#1,#2,#3){J.\ Phys.\ G  \issue(#1,#2,#3)} 
\newc{\PRDR}[3]{{Phys. Rev. D} {\bf #1}, Rapid  Communications, #2 (#3)}
%%%%%%%%%%%%%%%%%%%%%%%%%%%%%%%%%%%%%%%%%%%%%%%%%%%%%%%%%%%%%%%

%-----------------------------------------

%\vspace*{\fill}
%\vspace{-1.5in}
\begin{flushright}
{\tt IISER/HEP/07/10}
\end{flushright}
\begin{center}
{\Large \bf
The interplay between the charged Higgs and squark-gluino events at the 
LHC 
}
 \vglue 0.4cm
  Nabanita Bhattacharyya$^{(a)}$\footnote{nabanita@iiserkol.ac.in},
  Amitava Datta$^{(a)}$\footnote{adatta@iiserkol.ac.in},
  Monoranjan Guchait$^{(b)}$\footnote{guchait@tifr.res.in},
  Manas Maity$^{(c)}$\footnote{manas.maity@cern.ch} and
  Sujoy Poddar$^{(d)}$\footnote{sujoy$\_$phy@iiserkol.ac.in}
      \vglue 0.1cm
          {\it $^{(a)}$
          Indian Institute of Science Education and Research, Kolkata, \\
          Mohanpur Campus, PO: BCKV Campus Main Office,\\
          Mohanpur - 741252, India.\\
          \it $^{(b)}$
	   Department of High Energy Physics,
	   Tata Institute of Fundamental Research,\\
	   Homi Bhabha Road, Mumbai - 400005, India.\\
	   \it $^{(c)}$
	    Department of Physics,
	    Visva-Bharati,\\
	    Santiniketan - 731235, India.\\
          \it $^{(d)}$
          Netaji Nagar Day College,\\
          170/436, N.S.C. Bose Road, Kolkata - 700092, India.
          \\}
          \end{center}
          \vspace{.1cm}

\newpage
\begin{abstract}
{\noindent \normalsize}
In some extensions of the standard model with extended Higgs sectors,
events from  new particle production may pass the selection criteria for
Higgs search in different channels at the LHC - 14 TeV  and mimic 
Higgs 
signals. 
This intriguing possibility is illustrated by PYTHIA based simulations 
using several representative points in the parameter space of the 
minimal
supersymmetric standard model (MSSM) including a point in the minimal
supergravity model consistent with the Dark matter (DM) relic density data. 
Our simulations explore the interplay between the charged Higgs signal 
and typical
squark-gluino events. We argue that the standard selections like the one
based on the polarization properties of the $\tau$'s from charged Higgs
decay, though adequate for handling the SM background,  may not be very
efficient in the presence of SUSY backgrounds. We then
propose an alternative search strategy based on pure kinematics
which sufficiently controls both
the SM and the  MSSM backgrounds. For charged Higgs masses ($H^{\pm}$)
in the deep decoupling regime (600 GeV $\lsim m_{H^{\pm}} \lsim$ 800 GeV) this
method works well and extends the LHC reach close to 800 GeV for an
integrated luminosity of 30 $fb^{-1}$ with 
or
without the SUSY background. For a
lighter charged Higgs a judicious combination of the old selections  and 
some of the cuts proposed by us may disentangle the Higgs signal from the
squark-gluino backgrounds quite effectively.
\end{abstract}

%-------------------------------------------------------------------------

\section{Introduction}

The Standard Model (SM) has a single neutral Higgs scalar in the spectrum. 
In contrast many extension of the SM have multiple Higgs bosons, both 
charged and neutral.
Thus the discovery of at least two neutral Higgs bosons and / or a 
single charged scalar
would unambiguously indicate physics Beyond Standard Model (BSM). 
Search for these particles playing pivotal roles in electroweak 
symmetry breaking is, therefore, a high priority programme at 
the Large Hadron Collider (LHC) running at 14 TeV.

Almost all the models with an extended Higgs sector have  particles 
other than the quarks and leptons in the SM. They are expected to be 
produced at the LHC. The 
question that naturally arises is whether new particle production would 
create some final states which  may pass the selection criteria 
designed for Higgs search for taming down  the SM 
background only.
The "Higgs signal" at the end of the day then would be an admixture of  
genuine Higgs events and the "new physics"  backgrounds. It is therefore
important to improve the search strategy which eliminates  the SM as 
well as new 
backgrounds without paying too much price for the Higgs signal.

A case in point is models with supersymmetry (SUSY) \cite{susy}-
one of the most well 
motivated extension of the SM. The Minimal Supersymmetric Standard Model 
(MSSM) has two neutral scalars ($h$ and 
$H$), one neutral pseudo scalar ($A$) and two charged scalars 
($H^{\pm}$) in the Higgs sector \cite{susy,djouadi}. The 
prospective signals from these 
bosons at the LHC have been studied in great details 
\cite{djouadi,higgssearch,cmshiggs}. It 
needs to be emphasized 
that in addition to these bosons other superpartners of the SM particles, 
collectively called $\it {sparticles}$, will inevitably be produced, if 
they are within the kinematic reach of the LHC. Among the large variety of 
SUSY events there are 
quite a few sharing some or all features of one or more typical 
Higgs signals. The following intriguing question is then, can these  
events obfuscate the Higgs signal? The extreme 
case being events from the non-Higgs sector of the MSSM faking some 
MSSM Higgs signals even though the latter is too weak to be observed at 
the 
LHC- at least during the early runs. It is then important to improve the 
purity of the Higgs signal by 
introducing additional selection criteria which will suppress both 
SM and SUSY backgrounds. 
Now the issue is whether these additional cuts retains most of the 
Higgs signal without affecting  the discovery potential at the LHC.

In order to  illustrate the possible impact of SUSY events 
on 
non-standard Higgs signals, we simulate the charged Higgs 
signal in the MSSM  in the channel 
$g b \ra t H^{-} + c.c$, $H^- \ra \tau^- + \nu_{\tau}$
using the event generator PYTHIA \cite{pythia} along 
with squark-gluino events. The final state  of interest is 
one tagged $\tau$ jet, one tagged $b$ jet and 
missing transverse energy ($\met$). In addition
we require one reconstructed $W$ and one reconstructed top quark.
This signal has been studied by several 
groups~\cite{djouadi,higgssearch,cmshiggs,dp}. 
However, the selection criteria were designed to tame the SM 
background alone. These analyses are, therefore,
valid only if all sparticles 
are very heavy with negligible production cross sections at the LHC.  

It is clear that SUSY events with tagged $b$ and $\tau$ jets in the final 
state are potentially the most dangerous BSM background. 
Obfuscation of the Higgs signal will be even more serious if 
the events contain reconstructed $t$ quarks and $W$ bosons - either genuine 
or fake. To  demonstrate the interplay between the two sectors of the 
MSSM, we have chosen several points in the  parameter 
space (see Section 2). 
Some of them are almost tailor made for mimicking the
Higgs signal and we showcase our point
with these examples. We then change one or more key features of the 
parameter 
space considered to show that the interplay  happens over a much
wider parameter space.

We have mostly worked in the unconstrained MSSM without invoking any 
assumption 
regarding the boundary conditions at a high scale (say at $M_G$). 
However, we have also analyzed one mSUGRA scenario \cite{msugra}
with $\stau - \lspone$ coannihilation contributing to the 
DM relic density which turns out to be  consistent with WMAP data 
\cite{wmap}. This scenario also  provides a large  background to the 
charged Higgs signal.

In Section 3 we present a thorough analysis of the charged Higgs signal
as well as the SM and SUSY backgrounds and suggest kinematical cuts 
suitable for obtaining pure charged Higgs events while keeping the 
SM and the SUSY backgrounds at the minimal level. We also briefly
study the response of charged Higgs events to selection criteria for
squark-gluino search.   

The summary of the paper and the conclusions are in Section 4.
\section{The SUSY scenarios}

We begin with a MSSM parameter space with 
$m_{\tilde q} > m_{\tilde g}$, where the subscript 
$\tilde {q}$ refers to squarks of both L and 
R-types belonging to the first two generations. 
Thus gluino decays are dominated by the modes $\tilde g \ra f_3 \tilde 
f_3^*$ ,  where 
$f_3(\tilde f_3)$ denotes a third 
generation quark (squark) i.e, $t$ or $b$ ($\tilde t$ or $\tilde b$). 
Moreover $\tilde f_3$ is assumed to be significantly lighter than  
$\tilde q$. This is 
well motivated since the lighter mass eigenstates $\tilde t_1$ and 
$\tilde b_1$ are lighter than the $\tilde q$'s in general 
due to mixing in the third generation squark mass matrices. This 
assumption is especially appropriate for the $\tilde b_1$ since 
tan$\beta$ is large in scenarios with favourable production cross sections 
for the charged Higgs boson. Thus
squark-gluino events will surely contain quite a few hard taggable 
$b$-jets and genuine top quarks.

In the chosen  parameter space  the lighter chargino ($\chonepm$) and 
the second lightest neutralino ($\lsptwo$) dominantly decay into two body 
decay channels, $\chonem \to \stau_1^- \nu, \lsptwo \to \stau_1 \tau$,
since $\stau_1$, like $\tilde b_1$ is also naturally lighter than the other 
charged sleptons 
($\tilde e,\tilde \mu$) at large tan$\beta$. This leads to an 
abundance of taggable $\tau$ -jets in the final state.

Throughout this paper we compute the SUSY spectra from MSSM inputs  
using  {\tt SuSpect v2.3} \cite{suspect}, which takes 
into account radiative corrections to the sparticle masses.
We fix tan$\beta 
=40$ and $\mu$ = 500 GeV. In this section all masses, mass 
parameters and 
variables 
with dimensions of mass are in GeV. 
For simplicity we assume $M_1$ and 
$M_2$, the U(1) and SU(2) gaugino masses, follow the relation 
$M_1 =0.5 M_2$,
which is the typical expectation in models with a
unified gaugino mass in the electroweak sector at a high scale. 
But we assume $M_3$, the SU(3) gaugino mass to be a free parameter
not related either with $M_1$ or $M_2$ by any unification relation.
We shall, however, vary $M_3$ such that $M_3 > M_2$ and for some choices
$M_3$ also satisfies the unification condition.

It bears recall that in the decoupling regime ($m_A$ (the pseudo scalar 
Higgs boson mass) $ \gg m_Z$) the lighter Higgs scalar (h) has properties
identical to the SM Higgs boson. It is, therefore, imperative to 
discover at least
one more Higgs boson to establish an extended Higgs sector. The task 
becomes more challenging if the additional Higgs bosons are heavy.   
In view of this 
we consider several values of $m_A$ ranging 
from (500-800), which leads to approximately the same charged Higgs 
mass 
since, $m^2_{H^{\pm}} = m_A^2 + m_W^2$ (in the lowest order). 

In the first part of our analysis we do not employ any specific model.
Instead we choose general MSSM parameters constrained by general 
requirements like  absence of flavour changing neutral currents 
\cite{fcnc},
stability of the scalar potential \cite{ccb} etc. 
To begin with the following, masses are set to fixed values:
\br
m_{\tilde q_{L,R}} = 1000 + {\rm appropriate ~~D-term},
m_{\tilde g} = 560,
m_{\tilde e_{L,R}} = 303,
m_{\tilde \nu_{L}} = 293.  
\label{eq:one}
\er
From the input value of $M_1=150$ we obtain following masses:
\br
m_{\lspone} = 149, 
m_{\lsptwo} = 298,
m_{\chonep} = 298.
\label{eq:two}
\er
The masses for the 3rd generation sfermions are fixed by  
the following parameters:
\br
m_{t_L,b_L} = 600,
m_{b_R} = 500,
%m_{\tau_L,\tau_R} = 300,
m_{\tau_L} = 350,
m_{\tau_R} = 250,
A_t  = -900,
A_b  = -900,
A_{\tau}  = -500.
\label{eq:three}
\er
We first present our results for 
three choices of parameters SUSY I, II and III characterized by
$ m_{\tilde t_R}$=350,~400,~453 which along with
eq.~\ref{eq:three} lead to the spectra for the third generation 
sfermions in Table 1.
\begin{table}
\begin{center}\
\begin{tabular}{|c|c|c|c|c|c|c|}
\hline
Model & $m_{\tilde t_{1}}$ & $m_{\tilde t_{2}}$ & $m_{\tilde b_{1}}$
& $m_{\tilde b_{2}}$ & $m_{\tilde \tau_{1}}$ & $m_{\tilde \tau_{2}}$  \\
\hline
SUSY I& 306 & 677 & 500 & 630 & 215 & 378 \\
\hline
SUSY II& 353 & 683 & 500 & 630 & 215 & 378 \\
\hline
SUSY III& 397 & 690 & 500 & 630 & 215 & 378\\
\hline
\end{tabular}
\end{center}
\caption{ Masses of 3rd generation sparticles.}
\end{table}
In Table 2, we present the main decay channels of $\tilde g$, $\tilde 
t_1$ and 
$b_1$ squarks for the three scenarios in
Table 1. The mass of the lighter top squarks increases 
as we go from SUSY I - III. This drastically affects the 
gluino branching ratios (BRs) and 
number of genuine t - quarks in the final state. In all  cases, the 
dominant BRs of the  $\lsptwo$ and $\chonepm$ are practically 
fixed at 
$BR(\chonem \ra \stau_1^- \nu_{\tau}) = 0.91$ and
$BR(\lsptwo \ra \stau_1 \tau) = 0.91$. The first part of our 
analysis in the next section is based on SUSY I - SUSY III.   

\begin{table}[!htb]
\begin{center}\
\begin{tabular}{|c|c|c|c|}
\hline
Channels   & SUSY I & SUSY II & SUSY III\\
\hline
$\tilde g \ra \tilde t_{1} t$     & 0.80  & 0.61 & --\\
$\tilde g \ra \tilde b_{1} b$     &  0.18 & 0.38 & 1.0\\
$\tilde t_{1} \ra \chonep b$      &  1.0   & 0.40 & 0.47\\
$\tilde t_{1} \ra \lspone t$      &  --    & 0.60 &0.52\\
$\tilde b_{1} \ra \lspone b$      &  0.29  & 0.34 & 0.39\\
$\tilde b_{1} \ra \lsptwo b$      &  0.26  & 0.31 & 0.35\\
$\tilde b_{1} \ra \chonep t$      &  0.17  & 0.20 & 0.22\\
$\tilde b_{1} \ra \lstop W $      &  0.27  & 0.15 & 0.33\\
%$\tilde q_{L} \ra \tilde g q$    &  78.0 \\
%$\tilde q_{R} \ra \tilde g q$    &  94.4 \\$\chonepm \ra \stau_{1}
%$\chonepm \ra \stau_{1}  \nu_{\tau}$    &  89.5 & 89.4 & 89.4 \\
\hline
\end{tabular}
\end{center}
\caption{BR of gluinos and 3rd generation squarks.}
\end{table}

In the next phase of the analysis we fix the masses of the third 
generation sfermions as follows: 
$m_{\tilde b_{1}} = 755$,
$m_{\tilde b_{2}} = 980$,
$m_{\tilde t_{1}} = 751$,
$m_{\tilde t_{2}} = 1007$,
$m_{\tilde \tau_{1}} = 215$,
$m_{\tilde \tau_{2}} = 378$
and successively increase the gluino mass as follows,  $m_{\tilde g}$ = 
790, 950, 1020, 
1180 and 1345( SUSY IV - VIII) keeping
all other parameters  as in eq.~\ref{eq:one} and \ref{eq:two}. 
The first two choices (SUSY IV and V) correspond to $m_{\tilde g} < 
m_{\tilde q}$ and the 
gluino still decays dominantly into third generation squarks.  
However, the fraction of final states with genuine t quarks 
decreases. Hence, the  probability of the SUSY 
events faking the $H^{\pm}$ signal should also decrease in principle. In 
practice
, however, due to fake top reconstruction SUSY remains a potential
threat to the charged Higgs signal (see Section 3). 
For  $m_{\tilde g}=$ 1020 (SUSY VI) the squarks 
and gluinos are nearly degenerate. Finally for the 
last two choices $m_{\tilde g} > m_{\tilde q}$ and the gluino decays
to squark-antiquark or quark-antisquark  pairs belonging to the first 
two generations open up. This further reduces the presence of b and t in the 
final state. Apriorily the SUSY background is expected to be even smaller.

 In addition to the above parameter spaces 
we also analyze the following mSUGRA point consistent with the 
DM data~\cite{wmap},
\br 
m_0 =230,m_{1/2}=420, A_{0}=0,\tan\beta = 40,sign(\mu)> 0.
\label{eq:four}
\er
The resulting mass spectrum is calculated using {\tt SuSpect}~\cite{suspect} 
which gives for $m_t = 173$:
\br
m_{\tilde u_L} = 918,
m_{\tilde d_L} = 922,
m_{\tilde u_R} = 888,
m_{\tilde d_R} = 886,
m_{\tilde g} = 977,
m_{\tilde b_{1}} = 788, \nonumber \\
m_{\tilde b_{2}} = 849,
m_{\tilde t_{1}} = 694,
m_{\tilde t_{2}} = 866,
m_{\tilde e_{L}} = 365,
m_{\tilde e_{R}} = 280,
m_{\tilde \nu_{L}} = 356,
\nonumber \\
m_{\tilde \tau_{1}} = 182,
m_{\tilde \tau_{2}} = 370,
m_{\lspone} = 171,
m_{\lsptwo} = 323,
m_{\chonep} = 322
m_{H^{\pm}} = 500.
\er

In this scenario the relevant BRs are 
BR($\tilde g \ra t \tilde t_1$)=20 $\%$,
BR($\tilde g \ra b \tilde b_1$)= 26 $\%$,
BR($\chonem \ra \stau_1^- \nu_{\tau}$)= 95 $\%$ and
BR($\lsptwo \ra \stau_1 \tau $)= 94 $\%$.
The DM relic density  calculated by microOMEGAs (v 2.0) \cite{micromega}
yields $\Omega h^2 = 0.12$. 
The neutralino bulk annihilation \cite{bulk1,bulk} contributes 37$\%$ to 
the 
relic density
whereas $\lspone - \stau$ coannihilation \cite{bulk,staucoann}  
contributes 
63$\%$.
%($
%\%$ of Bulk, $\stau$ coannihilation ?).

In the following section we shall simulate   
the charged Higgs signal, the SM background and the SUSY events
corresponding to the above choices  of parameters.

\section{Simulation of the signal and the backgrounds} 
\label{result}

At the LHC the dominant contribution to single charged Higgs production
comes from the processes $g b \ra t H^{-} +$ c.c and 
$g g \ra t \bar{b} H^{-}$ + c.c.   
Both are related as they
stem from  gluon splitting ($g \ra b \bar{b}$) inside a proton. They
can, therefore, be regarded as two different approximation of the same
physical process \cite{alwall}.
%(ref 5, 6, 7).

For simulating charged Higgs production with an event generator at the
LHC, the process $g b \ra t H^{-}$ + c.c   along with parton showering 
is considered
if the additional b quark in the final state is not observable. The
initial b-quark is considered as one of the five massless partons in the
proton. In this approximation the $b\bar b$ pairs from the gluon
splitting belong to the region of the phase space where both are
collinear with the gluon. The resulting large logarithms due the
massless b-quarks can be consistently absorbed into the corresponding
parton density function (PDF). This gives a well defined leading order 
cross section.

We have simulated the leading order (LO) process,
\br
g b \ra t H^{-} + c.c 
\label{eq:five}
\er
with the top decaying hadronically: $t \to b q \bar q'$ and the charged 
Higgs 
into the  $\tau-\nu_{\tau}$ channel: $H^{-} \to \tau^- \nu_{\tau}$. This 
leads to a final state 
consisting a single $\tau$-jet, $b$-jet
accompanied by  missing energy due to the  
neutrinos and jets. We use 
PYTHIA \cite{pythia} to simulate the signal in eq. \ref{eq:five}. 
The cross sections are estimated  setting both 
renormalization and as well as factorization scale, 
$\mu_R = \mu_F = {\hat s}$ and using CTEQ5L PDFs
~\cite{cteql}. In the next to leading order (NLO) process
the $K$ factor for the signal process is $\approx$ 1.5~\cite{alwall}.

The dominant SM backgrounds are  due to the top pair production and
QCD events with jets  mis-tagged as $\tau$-jets.
We estimate these background along with the SUSY backgrounds
arising from squark-gluino events.
The LO cross section for $t \bar t$ is obtained
using CalcHEP \cite{calchep} (version 2.3.7).
We  require one top to decay
hadronically and the other into a $\tau$ and a
neutrino along with a $b$ quark.

The cross section for QCD has been computed by PYTHIA in two bins: 
(i)$400 < \hat {p_T} < 1000$ 
and (ii)$1000 <  \hat {p_T} < 2000$ GeV. The corresponding cross sections
are 2041 pb and 10 pb respectively. The contributions from other bins 
are negligible. Both the above cross sections are orders of magnitude 
larger than the signal cross sections (see Table 3)  and suitable 
kinematic selection will  be invoked to suppress them. 

Using ALPGEN \cite{wbg} we have also considered  the background from
from $W$ + $3$ jets , where all possible  jet combinations  including
a $b \bar b$ pair and a light jet have been considered.  
This background, however, is not very serious (see Table 3).

%In our 
%Tables we have presented $\sigma \times BR$ where $\sigma$ denotes leading
%order cross section for the signal and $BR$ is the branching fraction 
%of $H^{-} \ra \tau^{-} \nu_{\tau}$. 

In our simulation using PYTHIA we 
have taken into account the effects
of initial and final state radiation as well as 
fragmentation and hadronization.
A simple toy calorimeter simulation has been implemented with the 
following criteria:

\begin{itemize}

\item The calorimeter coverage is $\vert \eta \vert < 4.5$ with 
segmentation 
of $\Delta \eta \times \Delta \phi = 0.09 \times 0.09$ which 
resembles a generic LHC detector.

\item A fixed cone algorithm with $\Delta R$ = $\sqrt {\Delta\eta^2 + 
\Delta\phi^2}= 0.5 $ has been used for jet finding.

\item Jets are ordered in 
$E_T$ with $E^{jet}_{T,min} = 20$ GeV.
\end{itemize}

\underline{$b$- jet identification:}

We have tagged $b$-jets in our analysis by the following procedure.
%required only one tagged $b$-jet in our analysis.
A jet with $|\eta|< 2.5$ corresponding to the coverage of tracking detectors
matching with a $B$-hadron of decay length $> 0.9$ mm
has been marked $tagged$.
This criteria ensures that single $b$-jet tagging efficiency 
(i.e., the ratio of tagged $b$-jets and the number of taggable $b$-jets) 
$\epsilon_b \approx 0.5$ in $t \bar t$ events.
%Our $b$-jet tagging criteria is somewhat different from that in ref~
%\cite{guchait

\underline{$\tau$- jet identification:}

Taus are identified through their hadronic decays producing
narrow jets with 1 or 3 tracks pointing to the jets.
We have defined a narrow signal cone
of size $\Delta R_S= 0.1$ and an isolation cone of size $\Delta R_I= 0.4$
around the calorimetric jet axis. We then require 1 or 3 charged tracks
inside the signal cone with $|\eta_{track}|< 2.5$ and
$P_T > 3$ GeV for the hardest track.
We further require that there are no other charged tracks with $P_T > 1$ GeV
inside the isolation cone to ensure tracker isolation. 
%Thus we have an identified tau-jet(Cut 2).

\underline{Top quark reconstruction:}

We reconstruct one top quark following the procedure of \cite{top}
summarized below.
%For this reconstruction only hadronic decay channels of the top quark are
%considered. We first match the reconstructed jets with partonic
%quarks by defining a cone around the corresponding jet with $\Delta R =0.3$.
First we  compute the invariant mass of any two jets which are not 
tagged
as $b$-jets or $\tau$-jets, to reconstruct a candidate $W$. Further each
tagged $b$-jet is combined with a candidate $W$ to obtain a
candidate top quark. For each candidate a $\chi_{top}^2$ is defined:
\begin{equation}
\chi_{top}^2 = {({{m_W -m^{rec}_W} \over 15})}^2 + 
               {({{m_t - m^{rec}_t} \over 25})}^2
\end{equation}
\noindent
where $m_W = 80.42 $ GeV and $m_t = 173.1$ GeV are world averages of 
$W$-boson and top 
quark
masses and $m^{rec}_W$ and
$m^{rec}_t$ are reconstructed masses of $W$ -boson
and top quark candidates respectively.
We have  implemented $\chi^2$ minimization
procedure assuming a spread of 15 GeV and 25 GeV for the
reconstructed $W$ and top candidates respectively.
These numbers have been determined using 
Monte Carlo (MC) informations of jets originating from
the decays of $W$-boson and top quarks respectively.

Finally from MC information we have required minimum value of
$\chi^2$ for an acceptable event to be less than 8. This choice yields 
good efficiency for the reconstructed top in $t \bar t$ event.

The following selection criteria (SC) similar to those used in 
\cite{ritva}  have been used for rejection of the SM  background:

\begin{itemize}

\item We have required only one tagged $b$-jet (Cut 1).

\item We have asked for one identified $\tau$-jet (Cut 2). 

\item Events should have one detected $\tau$-jet with $E^{\tau-jet}_T > 100
$ GeV (Cut 3).

\item Events should have missing transverse energy $\etslash > 100$ GeV
(Cut 4).

\item  We require at least three jets in addition to one extra $\tau$ jet 
in the event (Cut 5).

\item We require one reconstructed top as described above (Cut 6).

\item We have also investigated the azimuthal opening angle
in the transverse plane between the $\tau$-jet and the $\etslash$ vector.
We have selected events with 
$\Delta \phi(\tau-jet,\etslash) > 60^{\circ}$ (Cut 7).

\end{itemize}

Table 3 shows the cumulative efficiencies of the Cuts 1-7  
for $m_{H^{\pm}} = 500, 600, 700, 800$ GeV
and $\tan\beta=$40 along with the $t\bar t$ and QCD backgrounds.
The  W + jets backgrounds are not shown as they are negligible.
Notice that in all cases except for the QCD background  
background, about 50\% of events  have only one tagged $b$-jet. 
Moreover, for all $m_{H^{\pm}}$, the overall efficiencies of the cuts for the 
signal  are roughly the same ($\sim  5-6\%$) , where 
as for the background
it is about 0.015\%. The $t \bar t$ background is mostly 
killed by the
strong cut on the  $E_T$ of the $\tau$-jets as the  $\tau$-jets 
originating from the charged Higgs in the decoupling regime are indeed 
much harder.
The final  cross section $\times$ efficiency ($\epsilon_1$) 
 for all types of events   after Cuts (1-7) are 
also presented (the appropriate BRs are included in
$\epsilon_1$). 
They lie in the range
0.8 - 4.2 fb. The corresponding  number for the total SM  background  is 
23.5 fb.

As is well-known,  the 
above cuts are inadequate to establish the charged Higgs signal,
even if the SUSY background is negligible.
For example, corresponding to  
 $m_{H^{\pm}}$= 500 (700) GeV which yields the largest 
(the third largest) signal cross
section the significance ( $S\over \sqrt{B}$) is 4.75 (1.58)
for $\lum_{int} = 30 \ifb$ . 

In Table 4 we present the cross sections and efficiencies for 
squark-gluino events computed by \cite{calchep}
for
the three model parameter spaces in Table 1, subject
to the same set of cuts. 
It 
should be borne in mind that 
the SUSY spectrum are the  same for the three sets except 
for $m_{\tilde t_1}$. We 
note that the 
final cross section of the SUSY events
is significant in all cases. In fact they are comparable to the $t 
\bar t$ background and even the weakest among them is larger than the 
signal for all $m_{H^{\pm}}$. 

As discussed in Section 2, the number of genuine top
quarks in SUSY III events is much smaller compared to the ones in SUSY I 
and SUSY II. Yet we find from Table 4 that the efficiency of getting 
one reconstructed top candidate is fairly high for all the SUSY scenarios under
consideration. This implies that
due to combinatorial backgrounds significant number of
fake top candidates are being reconstructed even in SUSY III.

We have also repeated the analysis
using the $t$-reconstruction prescription of
\cite{ritva}.  Our method yields better top reconstruction efficiency 
for the $ H^{\pm}$ signal. However, the procedure of \cite{ritva}
also yields sizable SUSY backgrounds via fake t-quarks in squark-gluino 
events.

The simultaneous presence of the Higgs and SUSY events leads to 
several interesting conclusions although some 
of them could  be misleading, as we shall see below. After the first set 
of Cuts (1-7) 
together they can
show up as a clear indication of  BSM physics standing over the SM 
background
although, as already noted, the Higgs events by themselves 
are not 
statistically significant. It is also important to note that this will
also disfavour a two Higgs doublet extension of the SM with a charged
Higgs of comparable mass and BRs. 

It follows from Table 3 and 4 that 
for $m_{H^{\pm}} = 500$ GeV and SUSY I 
the significance of
the combined BSM signal is 
7.5(13) for $\lum_{int} = 1 ~(10) \ifb$. 
If we consider the smallest contribution from the SUSY events  
(SUSY III), the significance is still 5.8 at 
$\lum_{int} = 10 \ifb$.
Thus the BSM physics can be established
at early stages of the LHC experiment 
although the final state will be an admixture
of Higgs and squark-gluino events.

For $m_{H^{\pm}} = 500 ~(800)$ GeV and SUSY I (III) 
Higgs events can be as large
as 27$\%$ (17 $\%$) of the number of SUSY events. 
However, there are cases where the
SUSY events simply dwarf the tiny presence of the Higgs. 
Consider, 
for example, SUSY II and $m_{H^{\pm}}$= 800 GeV, which 
corresponds to the weakest Higgs signal. 
Nevertheless, the significance of the combined 
signal is 6.5 for  $\lum_{int} = 10 \ifb$. This 
example clearly
demonstrates that even if the Higgs signal is negligible,
 SUSY events 
alone can masquerade as the charged Higgs signal.
Additional selection criteria are, therefore, called for  to disentangle 
the two types of events and confirm the Higgs signal.

It is known for a long time that the kinematic distribution of
the decay products of polarized $\tau$-leptons can be exploited
in new physics search \cite{bullock,manodp}. 
In conjunction with  the standard cuts (Cuts 1 - 7), $\tau$-polarization 
is
very effective in suppressing the SM backgrounds relative to the charged 
Higgs
signal (for both $m_{H^{\pm}} < m_t$ and $m_{H^{\pm}}  >  m_t$) 
in the $\tau-\nu_{\tau}$ 
channel \cite{bullock}. The SM background being suppressed is 
mainly due to the decay   
from $ W \rightarrow \tau \nu_{\tau}$. The main reason is that 
the polarization of the $\tau$'s in the charged Higgs decay ($P_{\tau}$= 
+1) and $W$ decay are opposite. As a result  the  
decay products of the $\tau$'s originating from the charged Higgs
have very different energy distribution than
their counterparts stemming from  W-decay .
In fact most of the subsequent analyses \cite{higgssearch,cmshiggs,  
ritva,sreerup}
have exploited this feature to improve the significance of the Higgs 
signal. In addition it also eliminates the QCD background very 
effectively.

However, cuts based on $\tau$-polarization may not be very efficient
in presence of SUSY backgrounds which is at the focus of interest
of this paper.
The  energetic $\tau$'s in the squark-gluino decay cascades,
which passes the selection criteria, mostly arise 
from the decays of $\stau_1$'s
($\stau_1 \ra \tau \lspone$). The 
polarization of these $\tau$'s may have a wide variety depending on 
the composition of $\stau_1$ and the 
lightest supersymmetric particle (LSP) 
\cite{nojiri}.
For example, if $\stau_1$ is dominantly a $\stau_R$ and the LSP is
bino like, the polarization of a $\tau$ in  SUSY cascades will be
identical to that of a $\tau$ arising from charged Higgs decay. 
This is true for all the parameter spaces considered by us.
In fact it 
has already been noted \cite{dpmanosusy}
that the polarization of the $\tau$'s in the squark-gluino decay 
cascades
are dominantly with $P_{\tau} \approx + 1$ in the mSUGRA type of model. This property  
enhances the 
SUSY signal in the interesting region of parameter space where LSP-
$\stau_1$ coannihilation \cite{staucoann} can generate the DM relic 
density of the universe. In general, however, the $\stau_1$ may be
an admixture of both $\stau_L$ and $\stau_R$. But even in this case a
significant fraction of the $\tau$'s will pass the selection criteria.   

The observations in the last paragraph can be justified
by the sparticle spectra and BRs presented in the last section.
For example in SUSY I  $\lstop \lstop^*$ pair production has 
a significant  cross section (3.3 pb) in SUSY I.
The $\lstop$ decays into $b \chonep$ with 100$\%$ BR.
The $\chonep$ dominantly decays into $\stau_{1}^{+}\nu_{\tau}$
(BR = 91 \%).
%Thus the
%major fraction of $\tau$s in this chain come from $\stau_1$
%which are very energetic. 
Thus the highly  energetic $\tau$'s in this decay chain  come from
$\stau_1$ decay.
A small fraction of the $\tau$s
come from the decay of $\chonep \ra \snu_{\tau} \tau^{+}$
%chi1->snutau+ tau 
followed by the invisible decay
of $\snu_{\tau}$ into $\nu_{\tau} \lspone$.
These $\tau$'s are rather soft because of the small mass difference 
between $\chonep$ and $\snu_{\tau}$ and fails to survive  the strong
cut on the $E_T$ of the $\tau$-jet.
%chargino and snutau.

Gluino pair production  also has a large cross section (4.6 pb).
From Table 2 it follows that
gluino decays dominantly into $\lstop t$ pairs.
As already discussed energetic $\tau$'s will come from $\stau_1$ 
decay cascade.
The $\tau$'s from W-decay are killed by the strong $E_T$ cut (Cut 3). 

Of course a small fraction (18$\%$) of the gluinos decay into
$\sbot b$. 
From Table 2 we can see $\sbot$ decays into $\lsptwo b$
with BR(26$\%$).
Here $\lsptwo$ decays into $\stau_{1} \tau$ with large BR(90$\%$).
Most of the primary $\tau$ in these decays are removed by Cut 3 
as already noted. The secondary $\tau$'s  from $\stau_1$
follows the polarization pattern as noted above.
Similarly the decay chain $\sbot 
\ra \chonem t$  yield energetic $\tau$'s with polarization properties as 
noted above. The other significant decay modes of $\sbot$ ($\lspone 
b$ and $\lstop W$) either yield no $\tau$'s or dominantly 
positively-polarized $\tau$'s.

In  SUSY I the squarks of first two generations decay
mainly into quark- gluino pairs. For example,
$\tilde u_L$ decays primarily into $\tilde g u$ with BR(78$\%$)
and $\chonep d$ with BR(13$\%$). As elaborated above the gluino
and chargino decays will mainly
lead to $\tau$'s with $P_{\tau}$ = + 1. 
The  $\tilde u_R$ decays primarily into $\tilde g u$ with BR(95$\%$)
and $\lspone u$ with BR(5$\%$). Thus
energetic $\tau$'s in the  squark decay chains will also
yield right handed polarized jets.

Applying similar chains of arguments to other SUSY scenarios (SUSY II 
-VIII) considered in this paper,
it is easy to see that in all cases the required
energetic $\tau$-jets 
dominantly come from $\stau_1$ decays and has $P_{\tau} \approx 1$. 

To roughly estimate the possible impact of the SUSY
background on  the Higgs signal, 
we note from Tables 1 and 2 of \cite{ritva}
that the significance of the signal for 
$m_{H^{\pm}} = 600$ GeV against a SM background 
of 0.22 $fb$ after Cuts 1-7 and the $\tau$-polarization cut is 5.6 at 
$\lum_{int} = 30 \ifb$.
If we assume that approximately all ( 50$\%$) of the weakest background 
from  SUSY III is retained after the $\tau$-polarization cut,
the significance reduces to 1.2 (1.6) for 30 $\ifb$.

In lieu of $\tau$-polarization we,therefore, 
add two more generic cuts which does not depend on the 
compositions of the sparticles. Instead they essentially depend on
the kinematics of the decay of a heavy Higgs. 
In order to motivate these cuts we 
present
in Fig. 1 the distributions of the $E_T$ of the $\tau$-jets 
and $\met$ (after Cuts 1-7) in the Higgs, SUSY   and 
SM events. Fig. 1 suggests that 
a stronger $\met$ cut removes the SM backgrounds efficiently while 
the stronger cut on the $\tau$-jet $E_T$ suppresses both SM and SUSY  
backgrounds,  while retaining bulk of the signal.

%%%%%%%%%%%%%%%%%%%%%%%%%%%%%%%%%%%%%%%%%%%%%%%%%%%

\begin{figure}[tb]
\begin{center}
\includegraphics[width=\textwidth]{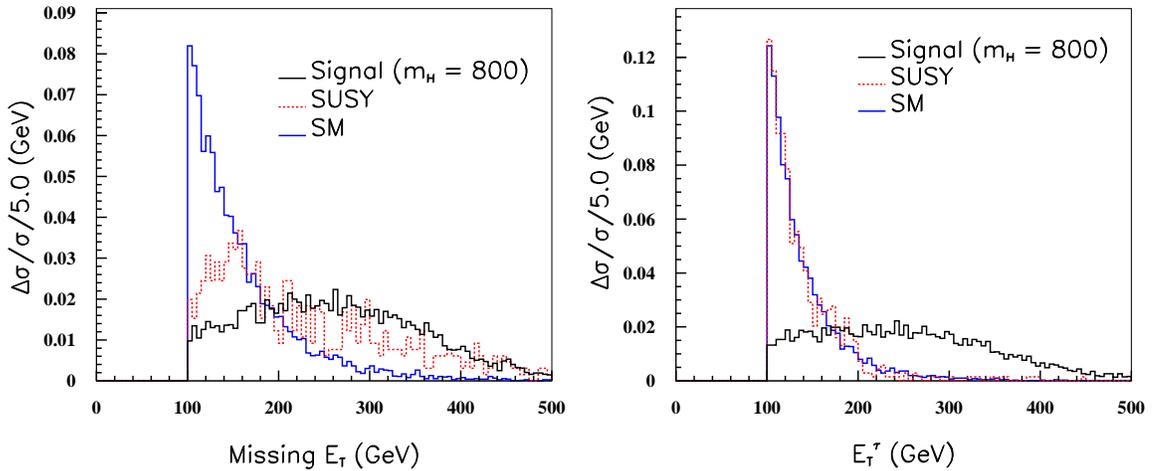}
\end{center}
\caption{
The distributions (normalised to unity) 
for signal (blue), SUSY (red) and SM (black) of  $\met$ (left)
and $E_T^{\tau -jet}$ (right) after selection Cuts 1-7.  Here
$m_{H^{\pm}} = 800 $GeV.
}
\end{figure}

%%%%%%%%%%%%%%%%%%%%%%%%%%%%%%%%%%%%%%%%%%%%%%%%%%%55

\begin{figure}[tb]
\begin{center}
\includegraphics[width=\textwidth]{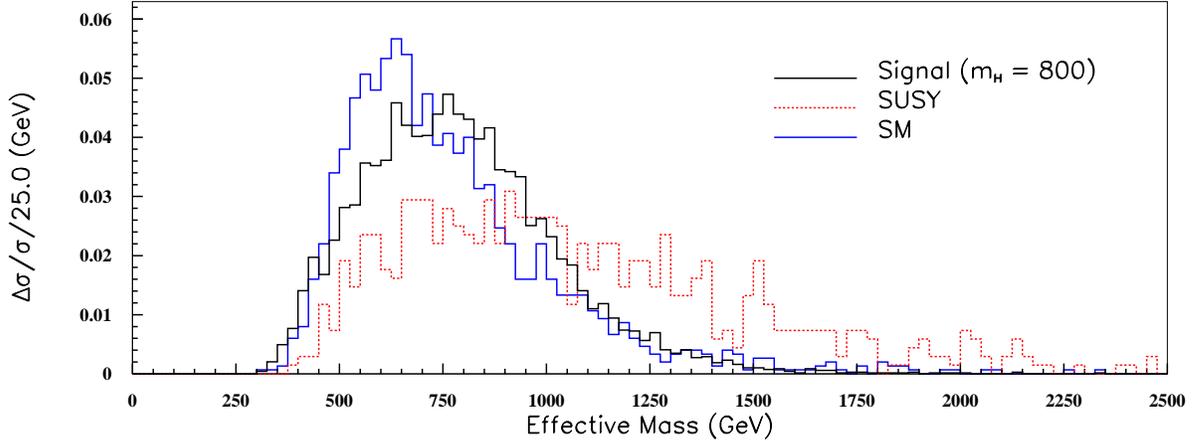}
\end{center}
\caption{
The distributions (normalised to unity) for signal (blue), SUSY (red)
and SM (black) of $M_{eff}$
after Cuts 1-7.
}
\end{figure}
%%%%%%%%%%%%%%%%%%%%%%%%%%%%%%%%%%%%%%%%%%%%%%%%%%%%%%%%%%%%%%%%%%%%%%%%%
%---------------------------------------------------------------------------
\begin{table}[!htb]
\begin{center}\
\begin{tabular}{|p{11mm}|p{27mm}|p{13mm}|p{13mm}|p{13mm}|p{13mm}|p{16mm}| p{16mm}| p{16mm}|}
%\begin{tabular}{|c|c|c|c|c|c|c|c|}
\hline
&    & \multicolumn{4}{|c|}{Signal ~$m_{H^{\pm}}$ (GeV)} & $t \bar t$ & QCD & $W +3j$\\
\cline{3-6}
 &    & 500          & 600          & 700          & 800 
   & &&\\
\hline
&$\sigma $ (pb)  &0.67   & 0.36& 0.20& 0.12 & 492 & 2042 & 46.65\\
\hline
&Selections &&&&&& &\\
\hline \hline

%NOE generated                                      & 100000 & 100000 & 
%100000 & 100000 &1000000 \\
Cut 1&1 $b-jet$                                     & 0.539 & 0.540 & 0.545 
& 0.546  &0.486 & 0.0745 & 0.266\\

%NOE with 1 hadronic $\tau$ leptons                 & 32671 & 32901 & 33287 
%& 33248&306531\\

Cut 2&1 $\tau-jet$                & 0.210 & 0.216 & 0.219 & 0.218 & 0.112 &
                                                   0.0017  & 0.005\\

Cut 3&$E_{T}^{\tau-jet} >100$ GeV    & 0.142 & 0.158 & 0.169 & 0.172 & 0.0086& 
         3.7$\times 10^{-5}$ & 4.6$\times 10^{-4}$ \\

Cut 4&$\etslash > 100$ GeV   & 0.118 & 0.137 & 0.152 & 0.157 & 0.0023 & 
          7.7$\times 10^{-5}$ &9.5$\times 10^{-5}$ \\

Cut 5&$N_{jet} \ge 3$(except $\tau$-jet) & 0.076  & 0.089  & 0.098  & 0.102 
& 0.0021& 6.1$\times 10^{-5}$  &8.2$\times 10^{-5}$  \\
Cut 6&1 reconstructed top & 0.054  & 0.061  & 0.066  & 0.069  & 0.0015 & 
                    9.0$\times 10^{-6}$ &2.1$\times 10^{-5}$ \\

Cut 7&$\Delta \phi(\tau-jet,\met)$$> 60^{\circ}$ & 0.051  & 0.058  & 0.064  
& 0.067  & 1.5$\times 10^{-4}$ & 6.0$\times 10^{-6}$ &- \\
\hline
&$\sigma \times \epsilon_1$ (fb) & 4.2 & 2.5 & 1.4 & 0.8 
& 11.2 & 12.3 &-\\
\hline

Cut 8&$E_{T}^{\tau-jet} >180$ GeV   & 0.025  & 0.036  & 0.045  & 0.050 & 
2.2$\times 10^{-5}$& - &-\\

Cut 9&$\etslash > 260$ GeV    & 0.0033   & 0.011  & 0.020  & 0.028 & 1.0$\times 10^{-6}$ &- &-\\

\hline
&$\sigma \times \epsilon_2$ (fb)& 0.27& 0.47& 0.45&0.34&
0.07 & - &-\\
%&$\sigma \times \epsilon_2$ (pb)& $2.7 \times 10^{-4}$& $4.7 \times 10^{-4}$ & 
%$4.5\times 10^{-4}$ & $3.4 \times 10^{-4}$ & \\
\hline
%Cut 10&$M_{eff} < 800$ GeV & 0.0018   & 0.0055   & 0.0076   & 0.0076& -- &-- &--%\\
Cut 10&$N_{jet}\le 6$        & 0.0028   & 0.0095   & 0.0181   & 0.0251& -- &-- & --\\
\hline
&$\sigma \times \epsilon_3$ (fb)& 0.24 & 0.41 & 0.40 & 0.30 & --& -- & --\\
\hline
\end{tabular}
\end{center}
   \caption{ The signal for different $m_{H^{\pm}}$ at $tan \beta = 40$ 
and the SM backgrounds
subjected to different selection criteria. 
}
\label{table1}
\end{table}
%---------------------------------------------------------------------------
The new kinematical selections  are:
\begin{itemize}
\item A more stingent cut on $\tau$-jet momentum, 
$E^{\tau-jet}_T > 180 $ GeV (Cut 8).
\item A stronger $\met$ cut,
$\met > 260 $ GeV (Cut 9).
\end{itemize}

We present the cross sections after Cuts 8 and 9 for the signal(Table 
3) along with the SM (Table 3) and the SUSY backgrounds(Table 4). In 
all cases 
$\epsilon_2$ is the combined efficiency of the Cuts 1-9.
The QCD background is eliminated. 
The $t \bar t$ background is $\sim$ 2.1  
for $\lum_{int} = 30 fb^{-1}$. The number of signal events for this
$\lum_{int}$ for $m_{H^{\pm}}$ = 500, 600, 700 and 800 GeV are 8.1, 14.1, 13.5 and 
10.2 respectively. In the absence of
SUSY backgrounds these cuts are adequate for establishing the
charged Higgs signal for $m_{H^{\pm}} > 500$ GeV.
For $m_{H^{\pm}} \lsim 500$ GeV higher luminosity may be required inspite 
of the larger production cross section, since
the Cuts 8 and 9 are less efficient.
We finally note that this generator level
analysis indicates that the  reach in $m_{H^{\pm}}$, in the absence of SUSY 
backgrounds,  is at least as good 
as that obtained in earlier works, if not better.

It also follows from Tables 3 and 4 that the new cuts improve the 
fraction 
of Higgs event
compared to the SUSY events in the combined BSM signal.
For example, corresponding to $m_{H^{\pm}} = 500 ~(800)$ GeV and SUSY I (III),
Higgs events can be as large
as 32$\%$ (180 $\%$) of the number of SUSY events.
The modest increase of the signal  
relative to the SUSY background for $m_{H^{\pm}} = 500$ GeV, 
suggest that this selection procedure
though very effective in the decoupling regime may not work for a 
lighter charged Higgs.       

However, even after the new cuts the Higgs signal is substantially
contaminated by the  squark gluino events in some cases. 
For example the significance of the Higgs signal
for $m_{H^{\pm}} = 600$ GeV and SUSY I (III) is 2.7 (5) at $\lum_{int} = 30 \ifb$
when the total background ( SM + SUSY) is taken into account.

To increase the  purity of the Higgs signal  we further require that

\begin{itemize}
\item $N_{jet} \le$ 6, where, $N_{jet}$ is the number of 
jets in an event (Cut 10). 
%\item $M_{eff}$ < 800 GeV, where $M_{eff} = |\met| +
%\Sigma_{i}|P_T^{l_i}| + \Sigma_{i}|P_T^{j_i}|$ ($l = e, \mu$).(Cut 10)
\end{itemize}
This cut reflects the fact
that the jet 
multiplicity in SUSY events is in general larger than
that in the Higgs signal. In Tables 3 and 4,
the total efficiency after Cuts 1-10 is  presented by  $\epsilon_3$.
Cut 10 practically eliminates the  $t \bar t$ events but
some SUSY backgrounds remain (see Tables 3 and 4).

We present in Table 5 the significance for 
different $m_{H^{\pm}}$ taking into account the largest
SUSY background (SUSY I) at $\lum_{int} = 30 \ifb$.
The signal is observable for $600 \lsim m_{H^{\pm}} \lsim 800$ GeV.
For $m_{H^{\pm}} = 500$ GeV the signal will be observable at
$\lum_{int} = 50 \ifb$. This again shows that 
this search strategy is more potent for $m_{H^{\pm}}$
in the deep decoupling region. We shall 
comment on lower Higgs masses in the following.

The number of jets in an event depend on the
parton showering model of PYTHIA. This has not been
tested in the LHC enviornment. We have, therefore,
attempted some other selection criteria and combination
thereof.

Since the signal is not expected to have many
isolated hard leptons (e,$\mu$),  one can veto 
events with a lepton
with $P_T^{e,\mu} > 15$ GeV and $\vert \eta \vert \le 2.4$ 
having $\Delta R_(jet,lepton) \ge 0.5$.
However this does not improve the significance of the signal,
see Table 5.

In Fig. 2 we present the distribution of
$M_{eff}$,
where $M_{eff} = |\met| +
\Sigma_{i}|P_T^{l_i}| + \Sigma_{i}|P_T^{j_i}|$ ($l = e, \mu$)
, for Higgs, SUSY   and
SM events. Fig. 2 suggests that a cut $M_{eff} < 800$ GeV 
would further suppress the SUSY background.
This affects the signal to a extent larger than Cut 10 
but practically
eliminates the SUSY background. The corresponding
number of background free signal 
events at $\lum_{int} = 30 \ifb$
have been presented
in Table 5 and does not look very attractive. 

A tighter cut $N_{jet} \le$ 5 gives a more
promising background free signal size.
This looks even more attractive if
the next to leading order cross section
with a $K =1.5$ is used.
Similar results have been obtained by
combining the lepton veto and $N_{jet} \le$ 6
criteria.

%---------------------------------------------------------------------------
\begin{table}[!htb]
\begin{center}\
\begin{tabular}{|p{12mm}|p{40mm}|p{17mm}|p{17mm}|p{17mm}|}
%\begin{tabular}{|c|c|c|c|}
\hline
&& SUSY I    &  SUSY II & SUSY III \\
\hline
&$\sigma $ (pb)                                & 12.2    &10.1  &8.9\\
\hline
%&$m_{\tilde t_R}$ & 350& 400 & 450\\
&$m_{\tilde t_1}$ (GeV) & 306& 353 & 398\\
\hline
&Selection Criteria &&&\\
\hline \hline
%&NOE generated       & 100000 & 100000 & 100000  \\

Cut 1&1 $b-jet$          &   0.32982  & 0.31205 & 0.24578  \\

Cut 2&1 $\tau-jet$       & 0.04088  & 0.03905 & 0.02912\\

Cut 3& $E_{T}^{\tau-jet} >100$ GeV                    &  0.00384   & 0.00379& 0.00381\\

Cut 4& $\etslash > 100$ GeV     &   0.00322   & 0.00294&  0.00323 \\  
Cut 5&$N_{jet} \ge 3$ (except $\tau$-jet)    &  0.00308   & 0.00281 &  0.00309\\

Cut 6 & 1 reconstructed top        & 0.00220   & 0.00204  &  0.00166  \\

Cut 7 &$\Delta \phi(\tau-jet,\met) > 60^{\circ}$     & 0.00129    & 9.1 $\times 10^{-4}$ 
   & 8.5 $\times 10^{-4}$\\
\hline
&$\sigma \times \epsilon_1$ (fb)      & 15.7   & 9.2  & 4.7\\
\hline

Cut 8 &$E_{T}^{\tau-jet} >180~ GeV$                  &  2.0$\times 10^{-4}$  &
                       1.9$\times 10^{-4}$ & 1.3$\times 10^{-4}$ \\
Cut 9 & $\etslash > 260~ GeV $                          & 7.0$\times 10^{-5}$
               & 8.0$\times 10^{-5}$  & 3.0$\times 10^{-5}$   \\  
\hline
&$\sigma \times \epsilon_2$ (fb)         & 0.85   & 0.80 & 0.19\\
\hline
%$S/\sqrt B $ at 100 $fb^{-1}$               &2.1 & 2.7 & 2.9 \\
%Cut 10 &$M_{eff} < 800 ~GeV$                            & -   & -   & - \\
Cut 10 &$N_{jet} \le 6$                            & 1.0$\times 10^{-5}$   
          & 1.0$\times 10^{-5}$   & 1.0$\times 10^{-5}$ \\
\hline
&$\sigma \times \epsilon_3$ (fb)       & 0.12   & 0.10 & 0.09\\
\hline
\end{tabular}
\end{center}
   \caption{The SUSY background for the sample points in 
Table 1.
}
\end{table}
%---------------------------------------------------------------------

\begin{table}[!htb]
\begin{center}
%\begin{tabular}{|c|c|c|c|c|}
\begin{tabular}{|p{25mm}|p{15mm}|p{15mm}|p{15mm}|p{15mm}|}
\hline
&\multicolumn{4}{|c|}{$m_{H^{\pm}}$ (GeV)}\\
\cline{2-5}
Cuts                 & 500  &  600  & 700 & 800 \\
\hline
&\multicolumn{4}{|c|}{$S/\sqrt B$}\\
\hline
$N_{jet} \le$ 6              & 4.0 & 6.5 & 6.3 & 5.0 \\
$N_{lep} =$0                 & 2.0 & 3.6 & 3.4 & 2.5\\
\hline
&\multicolumn{4}{|c|}{Background free number of signal events}\\
\hline
$M_{eff} < 800$ GeV       & 5 & 7  & 5  & 3 \\
$N_{jet} \le$ 5              & 6 & 10 & 10 & 7 \\
$N_{lep} =$0 and $N_{jet}\le$ 6 & 6 & 11 & 10 & 7 \\
\hline
\end{tabular}
\end{center}
\caption{ The significance of the Higgs signal or 
the number of background free signal events 
for different selection criteria at $\lum_{int} = 30 \ifb$. 
The background is
SUSY I.}

\end{table}

%---------------------------------------------------------------------

%Probe the $m^{max}_{\tilde g}$ that can spoil the $H^{\pm}$ signal.
%From Table 4 (SUSY IV, V) the observability of $H^{\pm}$ signal for 
%higher $m_{\tilde g}$.
%calculate $S/\sqrt B$ for 100 fb-1.

We next  increase $m_{\stau_1}$ to 275 GeV (
the earlier choice was 215 GeV) and re-calculate the SUSY background
keeping the other parameters in SUSY I fixed. There is a tread-off
between
the reduction in the number of   final states with $\tau$'s (e.g.,
$BR(\chonep \ra
\stau_1^{+} \nu_{\tau})$ now becomes  0.60 only) and the increased
efficiency of Cut 8. Finally, however, the SUSY background increases
substantially compared to SUSY I
( $\sigma \times \epsilon_3$ = 0.37 fb). Thus SUSY remains
a potentially dangerous background for a wide variety of $m_{\stau_1}$,
as long as the lighter electroweak gauginos decay with large BRs into
final states involving hard, taggable  $\tau$-jets.

So far our analyses were based on leading order cross sections
for the signal and the backgrounds. It is worthwhile to estimate the
possible impact of NLO cross sections. As already noted the K-factor for
the charged Higgs signal in the NLO is 1.5 \cite{alwall}. The
corresponding number for the $t\bar t$ is $\approx$ 1.6 \cite{ttnlo} .
Assuming that the efficiencies do not change drastically at the NLO ,
the significance of the Higgs signal with respect to the SM increases
slightly after Cuts 8 and 9 (compare with Table 3). The incorporation of
the K-factor for
squark-gluino production which, in this case is 1.3 - 1.4 \cite{sqnlo}
, may marginally increase  the final significance of the signal
vis a vis the SUSY background in Table 5.

We now present the SUSY backgrounds for different gluino masses 
(see Section 2) with a 
motivation
to probe the $m^{max}_{\tilde g}$ beyond which the SUSY
contamination in  the $H^{\pm}$ signal is negligible.
In Table 6, we display the raw cross sections and as well as 
effective cross sections ($\sigma \times \epsilon_i$).
The first two cases,  
SUSY IV and V stand for  $m_{\tilde g} < m_{\tilde q}$
where as last three scenarios, SUSY VI, VII, VIII correspond to   
$m_{\tilde g} \gsim m_{\tilde q}$.
Obviously 
the cross section and number of genuine top quarks in the final states 
decreases with increasing gluino mass 
since $\tilde g \to q \tilde q$ channel
open up with substantial branching fraction.
As a result the size of the SUSY background decreases
with increasing gluino mass. Yet in all cases
SUSY remains the dominant background after Cut 9.
After applying Cut 10 however, the SUSY background
is significantly reduced.
Again for $600 \lsim m_{H^{\pm}} \lsim 800$ GeV the
significance is $\gsim 5$ in all SUSY scenarios (IV -VIII).

%---------------------------------------------------------------------------

\begin{table}[!htb]
\begin{center}\
\begin{tabular}{|c|c|c|c|c|c|}
\hline
 & SUSY IV    &  SUSY V & SUSY VI & SUSY VII & SUSY VIII\\
$m_{\tilde g}$(GeV)& 790 & 950 & 1020 & 1180 & 1345 \\ 
$\sigma $(pb) & 5.6& 3.8 & 0.88 & 0.61 & 0.49  \\
\hline
$\sigma \times \epsilon_1$(fb)& 5.9  & 3.1  & 1.5 & 0.60 & 0.38 \\
$\sigma \times \epsilon_2$(fb)& 0.34 & 0.30  & 0.31 & 0.12 & 0.11 \\
$\sigma \times \epsilon_3$(fb)& 0.11 & 0.038 & 0.079 & 0.037 & 0.049\\
\hline
\end{tabular}
\end{center}
\caption{ The SUSY backgrounds for increasing $m_{\tilde g}$ (see
text for the details).
}
\end{table}

As already noted earlier the 
above search stratrgy may not be the best for $m_{H^{\pm}} \lsim 500$ GeV.
The stronger Cuts (8 and 9) may highly suppress the
signal.
We suggest that the earlier procedure 
\cite{djouadi,higgssearch,cmshiggs,dp}
based on the polarization of the $\tau$-jets
may be applied to suppress the SM background.
However, this could leave behind a large SUSY
background depending on the polarization property
of $\tau$'s appearing in the SUSY cascade decays.
A cut like  $N_{jet} \le$ 6 could then be implemented
for bringing the SUSY background under control.
It is already shown in Table 3 that one can 
implement this cut without paying too much
price for the signal.

Finally in Table 7 the result for a mSUGRA point motivated
by the DM data
point is presented.
Since $m_{H^{\pm}} = 500$ GeV in this case 
the significance of the signal
is rather modest as expected.
We find the significance at $\lum_{int} = 30 ~(100) \ifb$
is 3.4 (6.2).

%---------------------------------------------------------------------------

\begin{table}[!htb]
\begin{center}
\begin{tabular}{|c|c|}
\hline
 & mSUGRA point\\
$\sigma $ (pb)                                & 1.5     \\
\hline
$\sigma \times \epsilon_1$(fb)        & 1.35     \\
$\sigma \times \epsilon_2$(fb)        & 0.45    \\
$\sigma \times \epsilon_3$(fb)        & 0.15   \\
\hline
\end{tabular}
\end{center}
\caption{The SUSY background for a mSUGRA point motivated by DM
data (see Section 2). 
}

\end{table}
%>>>>>>>>>>>>>>>>>>>>>>>>>>>>>>>>>>>>>>>>>>>>>>>>>>>>>>>>>>>>>>>>n{table}[!htb]

%>>>>>>>>>>>>>>>>>>>>>>>>>>>>>>>>>>>>>>>>>>>>>>>>>>>>>>>>>>>>>>>>

%---------------------------------------------------------------------------

%\begin{figure}[tb]
%\begin{center}
%\includegraphics[width=\textwidth]{etmiss.eps}
%\end{center}
%\caption{}
%\end{figure}

\begin{table}[!htb]
\begin{center}\
%\begin{tabular}{|p{12mm}|p{40mm}|p{17mm}|p{17mm}|p{17mm}|p{17mm}|p{17mm}|}
\begin{tabular}{|c|c|c|c|c|}
\hline
& Selection Criteria           & $0 l$ & $1 l$ & $1 \tau + X$ \\
\hline
&Before all Cuts               & 0.732 &  0.240 & 0.124          \\
Cut 1$^{\prime}$ &$E_T^{jet1,jet2} > 150$ GeV   & 0.063 &  0.039 & 0.065          \\
Cut 2$^{\prime}$&$\met > 200$ GeV              & 0.084 &  0.110 & 0.197          \\
Cut 3$^{\prime}$ &$M_{eff} > 1000$ GeV          & 0.437 &  0.471 & 0.350          \\
Cut 4$^{\prime}$ &Transverse sphericity $> 0.2$ & 0.414 &  0.541 & 0.428          \\
\hline
&$\sigma \times \epsilon_4$ (fb)&2.03  & 0.75 & 0.69\\
\hline 
\end{tabular}
\end{center}
\caption{The effective cross sections
of $0 l$ , $1 l$ and $1 \tau + X$ events
subjected to
Cuts(1$^\prime$ - 4$^\prime$) for $m_{H^{\pm}} = 300$ GeV.
}
\end{table}

We now address the inverse problem. Namely,  how the charged Higgs 
events  may affect the canonical 
$m$-leptons + $n$-jets + $\etslash$ signatures
of  squark-gluino production. We have restricted ourselves to  
$m =0,~1$. 

In our simulation
leptons $(l=e,\mu)$ are selected with P$\mathrm{_T \ge 30}$ GeV
and $\vert\eta \vert \le 2.5$. For lepton-jet isolation
we require $\Delta R(l,j) > 0.5$. The detection efficiency of the leptons 
are assumed to be $ 100 \%$ for simplicity.

We have also looked into final states of the type $1\tau + X$
where $X$
includes two or more hard jets but no $e$ or $\mu$ or tagged $\tau$. 
Tagging of
$\tau$ jets are implemented according to the following procedure.

Only hadronic $\tau$ decays
are selected. The $\tau$-jets with $\eta <$3.0 are then divided into
several $E_T$ bins. A $\tau$-jet  in any $E_T$ bin is then treated as tagged
or
untagged according to the efficiency ($ \epsilon_{\tau}$) given in
\cite{cms} Fig. 12.9 for a particular bin.

We have implemented following selection criteria 
(see Chapter 13 of \cite{cmshiggs}):
\begin{itemize}

\item We select events with at least two jets having P$\mathrm{_T} > 150$
GeV (Cut 1$^{\prime}$).

\item Events with missing energy ($\etslash) > 200$ GeV are selected (Cut 2$^{\prime}$).

\item Events with $M_{eff} > 1000$ GeV are selected,
where $M_{eff}= |\met| + \Sigma_{i}|P_T^{l_i}| + \Sigma_{i}|P_T^{j_i}|$
($l = e,\mu$ ) (Cut 3$^{\prime}$).

\item Only events with jets having S$\mathrm{_T} > 0.2$, where
S$\mathrm{_T}$ is a standard function of the eigenvalues of the
transverse sphericity tensor, are accepted (Cut 4$^{\prime}$).
\end{itemize}

For $m_{H^{\pm}} = 300$ GeV the LO production cross section
is 1.29 pb.
In Table 8 we  present the effective cross sections 
($\sigma \times \epsilon_4$) of 
$0 l$ , $1 l$ and $1 \tau + X $ events
after implementing the
Cuts(1$^\prime$ - 4$^\prime$) for $m_{H^{\pm}} = 300$ GeV.
Here $H^{\pm}$ is allowed to decay in all possible modes.
We find that the number of Higgs induced 
events for $\lum_{int} = 10 \ifb$ are 
20.3, 7.5 and 6.9 respectively.
These numbers though numerically significant,
are unlikely to affect
the result of SUSY search. 
Our earlier analyses clearly suggest that the 
strong $M_{eff}$ (Cut 3$^{\prime}$)
protects the squark-gluino signals from
contamination due to Higgs induced events.
The number of events for $0 l$ for $m_{H^{\pm}} = 400, 500$ GeV
at $\lum_{int} = 10 fb^{-1}$ are 22 and 18 respectively.

\section{Conclusions}

In many models with extended Higgs sectors there are varities of new 
particles. It is then pertinent to ask how many events stemming 
purely from new 
particle production, can pass the slection criteria for Higgs search at 
the LHC, usually designed to remove the SM background alone. Such events 
(``the new physics" background) if sizable in numbers will 
obviously obfuscate the Higgs signal. Additional kinematic selection 
must, therefore, be carefully designed to suppress the new backgrounds 
while the bulk of the Higgs signal is retained. 

It should also be 
stressed that before these additional cuts the Higgs and the new physics 
events together may stand above the SM background and reveal BSM physics 
at early stages of the LHC experiment, using selection criteria quite 
different from the cannonical search strategies for new physics.

We illustrate this very generic possibility by charged Higgs ($H^{\pm}$) 
search in the MSSM taking into account the SUSY backgrounds from squark 
gluino events. In the decoupling regime ($m_A \gg m_Z$) the lighter 
scalar ($h$) mimics the SM Higgs boson. The first step for establishing 
the extented Higgs sector of the MSSM would, therefore, be to discover 
another Higgs boson. The early discovery of the charged Higgs can 
adequately serve this 
purpose. The larger the $m_H^{\pm}$, the more challenging the discovery 
would be. Our main analysis is focussed on charged higgs search in the 
deep decoupling regime with $500 \leq m_{H^{\pm}}\leq 800$ GeV). We 
also comment on the prospect of lighter charged Higgs search in the 
presence of SUSY backgrounds.  

There already exists in the literature several strategies to tame the SM 
background to the Higgs signal \cite{higgssearch,cmshiggs,ritva}. A variety of 
points in the MSSM parameter space with different characteristics 
including a mSUGRA point consistent with the observed DM relic density 
are considered (see Section 2) for sampling the SUSY backgrounds. 
The background events  have a large number of taggable 
$\tau$-jets, b-jets and either genuine or fake reconstructed top quarks.

At the first stage of our analysis, we implement Cuts 1-7 (see Section 
3, Table 3) following \cite{ritva}. They suppress the SM backgrounds 
quite a bit, but- as is well known- are not enough to establish a 
statistically significant Higgs signal  without additional cuts. If 
SUSY backgrounds are also 
present, the Higgs signal will be swamped by the combined background ( 
see $\sigma \times \epsilon_1$ in Tables 3, 4, 6 and 7). On the other 
hand the combined Higgs and squark-gluino events in the scenarios 
considered by us would stand above the SM background and establish BSM 
physics using selection criteria quite different from that typically 
implemented for SUSY search alone. In fact the outcome of this analysis 
would also disfavour the two Higgs doublet extension of the SM with comparable 
charged Higgs mass and BRs. Yet the new physics events will be an 
admixture of Higgs and squark-gluino events with no clear evidence of 
the Higgs signal.

In the standard analyses ignoring the SUSY background,  the second 
stage consists of additional cuts 
which further suppress the SM background and establish the Higgs signal 
at a higher level of confidence. The polarization properties of the 
$\tau$'s stemming from the charged Higgs decay are utilized in many 
analyses (see \cite{manodp} and references there in). However, as
argued in Section 3, the selections based on
$\tau$-polarization may not be very efficient if SUSY backgrounds 
are present.

In lieu of $\tau$-polarization we select a more generic set of cuts
(Cut 8-9, see Table 3) which depends on kinematics rather than
on the compositions of the sparticles. In the absence of SUSY 
backgrounds, the estimated reach for $m_{H^{\pm}}$  on  the basis of this 
generator level analysis would be close to 800 GeV for $\lum_{int} = 30
fb^{-1}$. A full simulation based on this alternative strategy would be
welcome.

The above cuts  not only 
bring the SM backgrounds
further down but also enrich the fraction of the Higgs induced events 
in the 
surviving sample ( see $\sigma \times \epsilon_2$ in Tables 3, 4, 6 and 
7). But  the SUSY backgrounds may still be too large for
an unambiguous Higgs discovery as is illustrated by the above examples. 
In fact even if the charged Higgs signal is too low, the SUSY background
can fake it by standing above the SM background.

For selectively suppressing the SUSY events one has several 
alternatives. In Tables 3, 4, 6 and 7  we display the effect of a cut on 
jet multiplicity (Cut 10) based on the fact that this 
number in  a typical SUSY event is generally larger than that 
in the
Higgs signal. After this cut an almost pure Higgs sample is left behind
( see $\sigma \times \epsilon_3$ in Tables 3, 4, 6 and 7).

In contrast if the size of the event sample remain practically unaltered
after Cut 10, it would imply that no significant SUSY background was 
left behind after Cut 9.  This by itself may be indicative of the 
nature of the SUSY parameter space. For example, scenarios with 
electroweak gauginos decaying via two body modes (see Section 2) into 
final states with energetic $\tau$-jets will be disfavoured. 

The significance of the Higgs signal with respect to the scenario with 
the largest SUSY background (SUSY I )is displayed in Table 5. For $600 
\leq m_{H^{\pm}} \leq 800$ GeV, the significance is $\geq 5$ for $\lum_{int} = 30 
fb^{-1}$. For $m_{H^{\pm}} = 500$ GeV, $\lum_{int} \sim 50 fb^{-1}$ may be required. 

We have also   
discussed several alternatives for finally suppressing the 
SUSY background. (see Table 5). 
The corresponding significance ( or the number of signal events if 
the background is zero) are also displayed in Table 5. It seems that  
the efficiency of the 
cut on jet 
multiplicity 
($N_{jet} \leq 6$ ) is marginally better. However, some selection 
criterion makes the signal background free according to our generator 
level calculation. If this conclusion survives a full simulation, then  
NLO cross 
sections would predict a signal size larger by a factor of 1.5 and 
the Higgs discovery could be made at a higher level of confidence.   
 
It follows from Table 5 that the above  strategy does not look very 
promising   
for  lighter Higgs bosons ($m_{H^{\pm}} \leq  500$ GeV). The main point is that for 
lower Higgs mass the
$\tau$-jets are not sufficiently hard to pass the stiff $E_T$ cut
(Cut 8). In such cases - after the standard cuts - one may implement 
the $\tau$-polarization cut which is indeed effective in removing the SM 
backgrounds. Finally Cut 10 or some of its alternatives may be used to 
further suppress the SUSY backgrounds to give a pure Higgs sample.

	In conclusion we reiterate  that the interplay between the 
Higgs and squark- gluino events  could be relevant in principle 
for all  Higgs (charged or neutral) search channels. More care in 
designing the selection procedure- keeping in mind the possible 
interplay 
between Higgs and squark-gluino events - for all the Higgs 
search channels is, therefore, called for. Since the key issue of 
electroweak symmetry breaking hinges on the Higgs sector this additional 
attention is indeed justified. 
   
%---------------------------------------------------------------------------
{\bf Acknowledgment}:
NB
would like to thank the Council of Scientific and
Industrial Research, Govt. of India for financial support.
MM acknowledges support from Department of Science and Technology,
India (grant SR/MF/PS-03/2009-VB-I).

%---------------------------------------------------------------------------

\end{document}
%-------------------------------- ---------------------------------------